\documentclass{article}
\usepackage[utf8]{inputenc} 
\usepackage{amsmath}
\usepackage{graphicx} 
\usepackage{booktabs}  
\usepackage{multirow}
\usepackage[a4paper,left=3cm,right=3cm,top=3cm,bottom=3cm]{geometry}

\usepackage{wrapfig}
\usepackage{float} 
\usepackage{color}
\usepackage{caption}
\usepackage{subcaption}
\usepackage{hyperref}
\usepackage{centernot}
\usepackage{comment}
\usepackage{cite}
\hypersetup{
  colorlinks   = true,    
  urlcolor     = blue,   
  linkcolor    = blue,    
  citecolor    = black     
}

\newcommand{\KaTie}{{\sc Ka\hspace{-0.2ex}Tie}}
\DeclareUnicodeCharacter{2212}{-}
\DeclareUnicodeCharacter{03B1}{\ensuremath{\alpha}}

\title{TEEC and azimuthal correlations in forward $\gamma$-hadron production in proton-proton and proton-lead collisions at LHC}
\author{
Ishita Ganguli$^a$ \thanks{Corresponding author. E-mail: \texttt{ganguli@agh.edu.pl}},
Piotr Kotko$^a$ \thanks{E-mail: \texttt{piotr.kotko@fis.agh.edu.pl}}
\\ \\
$^a${\it AGH University Of Krakow,}\\ 
{\it Faculty of Physics and Applied Computer Science,} \\ 
{\it al. Mickiewicza 30, 30-059 Krak\'ow, Poland} 
}

\date{}

\begin{document}

\maketitle

\begin{abstract}
    We investigate photon-hadron Transverse Energy-Energy Correlation (TEEC) and azimuthal correlations in the forward rapidity region for proton-proton and proton-lead collisions within the small $x$ saturation formalism based on the Color Glass Condensate (CGC), supplemented by the initial-state and final-state parton shower simulation and hadronization using CASCADE. The study was performed in light of the upcoming FoCal detector of ALICE and the ATLAS detector upgrades with the extended pseudorapidity coverage. We predict a significant difference, at hadron-level, between the normalized TEECs in proton-proton and proton-lead collisions. Having a direct access to partonic subprocesses we also directly investigate the impact of the parton shower and hadronization on the TEEC. The ordinary azimuthal $\gamma$-hadron correlations show substantial suppression for proton-lead collision, as predicted earlier at parton-level.
\end{abstract}

\section{Introduction}

Event shape observables, such as thrust, have long been used in precision studies in Quantum Chromodynamics (QCD), see eg. \cite{L3:1992btq, DECAMP1991479, Banfi:2010xy}. They are defined to be infrared and collinear safe, meaning they are designed to be robust against soft gluon emissions and collinear splittings. By quantifying complex multi-particle final states with the use of simple scalar quantities, event shape variables allow for precise comparisons between experimental measurements and perturbative QCD calculations. Event shape observables in $e^{+} e^{-}$ collisions have been studied up to next-to-next-to-leading order at fixed order \cite{Ridder_2007, Weinzierl_2008, Weinzierl_2009}, and at next-to-next-to-next-to-leading logarithm \cite{Becher_2008, Abbate_2011, Chien_2010}. Energy-Energy correlators (EEC), first introduced in \cite{Basham:1978bw,Basham:1978zq}, are more differential variants of the C-parameter event shape variable and measure the correlations of energy flow along directions on celestial sphere see e.g., \cite{ACTON1992547, PhysRevD.31.2724, AKRAWY1990159} for experimental determination and \cite{Neill:2022lqx,Moult:2025nhu} for more comprehensive reviews and summary of recent computations. Here, let us mention, that  despite the relative simplicity of the observable, the full analytical result at NLO level in QCD was computed not that long ago \cite{Dixon:2018qgp}. In parallel, let us note that EEC are interesting observables to apply holography in order to study EEC's non-perturbative aspects  \cite{Hofman:2008ar} (see also \cite{Moult:2025nhu} and references therein). Transverse Energy-Energy Correlators (TEEC) \cite{ALI1984447} are a natural extension of the EEC for hadron colliders. TEEC, by definition, suppresses contributions from small transverse energy hadrons. Final states in TEEC can be jets or hadrons. In the latter case, it gives quite a direct access to underlying hard partonic process, due to the suppression of soft radiation (see e.g. \cite{tao2020} in the context of DIS). Jet TEEC was measured at LHC in the context of strong coupling studies \cite{ATLAS:2015yaa, ATLAS:2017qir, ATLAS:2023tgo}. However, in general, sound jet reconstruction requires significant $p_T$, which is an unwanted feature in \emph{gluon saturation} searches that should occur at small $x\sim p_T/\sqrt{s}$. Therefore, in the following work we shall focus on TEEC defined for hadrons.

The process we are looking into is the forward production of photons and hadrons, 
\begin{equation}
    p (P_A) + A (P_B) \rightarrow \gamma (p_{\gamma}) + h (p_h) \,,
\end{equation}
where $A$ is either a proton or a lead nucleus. At the perturbative level, a single colored state makes this observable a favorable starting point for gluon saturation studies in the near-future high luminosity LHC forward calorimeters \cite{dEnterria:2025jgm}. Gluon saturation phenomenon was predicted in QCD long time ago \cite{GRIBOV19831,Mueller:1985wy} (see also the book \cite{Kovchegov:2012mbw}), and later developed mainly through the Color Glass Condensate (CGC) effective theory (see e.g.~\cite{Gelis:2010nm} for a review). 
The computations of scattering processes in CGC are incomparably more involved than in the collinear factorization. There are a multitude of reasons for that, in particular, the CGC computations include multiple interactions and thus higher twist corrections. Production of photons has been computed in CGC at LO in \cite{Rezaeian:2012wa, Jalilian-Marian:2012wwi} with many follow-up computations for proton-proton and proton-nucleus collisions \cite{Ducloue:2017kkq, Goncalves:2020tvh, SampaiodosSantos:2020raq, Golec-Biernat:2020cah, Santos:2020kgv, Benic:2022ixp, Lima:2023dqw, Ganguli:2023joy, Adhya:2024pgm} as well as electron-nucleus collisions \cite{Roy:2019hwr, Roy:2018jxq, Kolbe:2020tlq}.

On experimental side, $\gamma$-hadron correlations in proton-proton and proton-lead collisions have been measured by ALICE \cite{ALICE:2020atx, ALICE:2025bnc}, while $\gamma$-jet correlations have been studied by ATLAS \cite{ATLAS:2012ar, ATLAS:2017xqp} for proton-proton collisions only. 
Although some of these studies do indicate the presence of nuclear initial state effects, these measurements were performed in lower rapidity range and were not designed to search for gluon saturation effects.
While a direct observation of gluon saturation remains challenging, several experimental signatures serve as indirect evidence \cite{BRAHMS:2004xry, Aaron:2009aa, STAR:2021fgw}; still, there are other mechanisms that can potentially be at play, such as the so-called leading twist nuclear shadowing  \cite{Frankfurt:2011cs} or parton energy loss. 
Discriminating between different mechanisms requires a selection of observables and broad range of kinematics.

In view of the above, the purpose of this work is to compute both TEEC and photon-hadron azimuthal correlations, in the kinematics of the near-future ALICE and ATLAS upgrades. FoCal, or Forward Calorimeter \cite{ALICE-PUBLIC-2023-001, ALICE:2024jtt} is a part of the ALICE detector in LHC and is planned to cover a very forward pseudorapidity window: $3.8 < \eta < 5.1$. On the other hand, the ATLAS calorimeter is proposed to have an increased pseudorapidity coverage of $|\eta|$ up to 4 \cite{Solovyanov:2024gri}. The distinctive feature of our computation is inclusion of the full parton shower and hadronization, on top of the CGC setup (in the form of the hybrid $k_T$-factorization that convolutes the dipole TMD gluon distribution with kinematic constraints and DGLAP correction with off-shell gauge invariant leading order amplitude). The work is a significant extension of \cite{Ganguli:2023joy} and provides a parallel study to \cite{Kang:2025vjk} from the Monte Carlo perspective.

As the final state photon can either be a direct product of the hard interaction, or can be created later during the shower, such a hadron-level computation has many challenges. At present there is no \emph{general purpose} event generator that does the parton-level computation within the CGC framework, followed by the parton showers and hadronization. The most serious advancement is the nonlinear parton shower based on Gribov-Levin-Ryskin equation \cite{Shi:2022hee}, also with the kinematic constraint \cite{Shi:2023ejp}. There are also other generators that include certain small-$x$ effects or Pomeron self-interactions (see e.g. \cite{Engel:1994vs, Roesler:2000he, Werner:2005jf, Werner:2023zvo, Fletcher:1994bd, Ahn:2009wx, Kalmykov:1997te, Ostapchenko:2010vb}). However, in our computation we would like to adhere to TMD-like factorization directly based on the CGC framework.  While the leading order partonic cross section is straightforward to compute using e.g. \KaTie\, Monte Carlo \cite{katie}, there does not exist a Monte Carlo generator that can perform a shower with gluon recombination, consistent with the Balitsky-Kovchegov (BK) nonlinear evolution equation \cite{Balitsky:1995ub, Kovchegov:1999yj} for TMD gluon distribution, not mentioning the full B-JIMWLK equation \cite{Balitsky:1995ub, Jalilian-Marian:1997qno, Kovner:2000pt, Kovner:1999bj, Weigert:2000gi, Iancu:2000hn, Ferreiro:2001qy}. There is however the CASCADE Monte Carlo \cite{Baranov_2021}, which is capable of turning the information contained in the TMD gluon distribution into a cascade of partons (see also \cite{Bury:2017jxo}). Such an approach has already been used in the past, see \cite{Bury:2017xwd, VanHaevermaet:2020rro} and this will be our choice for the present study.

The work is organized as follows. In the next section we explain the formalism used to perform our computations and define the observables. In Section~\ref{sec:Results} we list details of the kinematic cuts and discuss the results. Finally, in~\ref{sec:Summary} we summarize the paper.

%%%%%%%%%%%%%%%%%%%%%%%%%%%%%%%%%%%%%%%%%%%%%%%%%%%%%%%%%%%%%

\section{Framework}
\label{sec:Framework}

Our computation is a two-stage procedure.
\begin{enumerate}
    \item 
First, we generate the Monte Carlo samples for the partonic processes $q (\Bar{q})g^*\rightarrow q(\Bar{q})\gamma$, where $g^*$ represents an off-shell gluon, that, in addition to the transverse momentum, carries only longitudinal momentum parallel to the mother hadron (nucleus), but no longitudinal momentum of the other hadron. Such an approximation is dictated by the high energy limit in the spirit of the high energy factorization \cite{Catani:1990eg, Collins:1991ty}, or, actually, the so-called hybrid factorization \cite{Dumitru:2005gt, Deak:2009xt}). Indeed, the forward rapidity requirement on the final states translates to the asymmetry of the fractions $x_A$, $x_B$ of the longitudinal hadron momenta; more precisely
\begin{equation*}
    x_{A} = \sum_{i} \frac{|\vec{k_{T\, i}}|}{\sqrt{s}} e^{-\eta_i} , \,\, x_{B} = \sum_{i} \frac{|\vec{k_{T\, i}}|}{\sqrt{s}} e^{\eta_i}\, .
\end{equation*}
Therefore, for the large positive rapidity of the final states the $x_A$ fraction will be small, while $x_B$ moderate, enabling the approximation in which the hadronic state probed at $x_B$ is ``dilute" and evolves according to the ordinary DGLAP evolution, while the one probed at $x_A$ is ``dense", demanding the use of the CGC framework. However, if the transverse momenta of the final states are sufficiently large (but not too large -- to still permit the use of the small-$x$ approach), the CGC framework can be reformulated into the TMD factorization framework \cite{Dominguez:2011wm,Kotko:2015ura,Altinoluk:2019fui}.
The cross section for producing the $q(\Bar{q})\gamma$ final state is computed via the following factorization prescription: 
\begin{equation}
    d\sigma_{\text{p}A\rightarrow  \gamma q (\Bar{q})} = \int\! dx_A\, dx_B \int\!\! d^2k_T\, f_{q (\Bar{q})/B}(x_B;\mu)\,\mathcal{F}_{g}\left(x_A,k_T;\mu \right)   d\sigma_{q (\Bar{q})g^*\rightarrow \gamma q (\Bar{q})}(x_A,x_B,k_T;\mu) \,,
\end{equation}
where $f_{q (\Bar{q})/B}$ is the collinear PDF corresponding to the quark or anti-quark originating from the proton projectile $B$, $\mathcal{F}_{g}$ is the \emph{dipole} gluon TMD PDF corresponding to the gluon coming from the target $A$ (we skip the operator definition here -- see eg. \cite{Ganguli:2023joy}). The partonic cross-section $d\sigma_{q (\Bar{q})g^*\rightarrow \gamma q (\Bar{q})}$ is calculated in a gauge invariant way with one off-shell gluon carrying four-momentum $k_A^{\mu} = x_A P_A^{\mu} + k_T^{\mu}$. The initial-state quark/anti-quark in this approximation does not carry any transverse momentum component; its four-momentum is $k_B^{\mu} = x_B P_B^{\mu}$.

The partonic Monte Carlo samples were generated in \KaTie\, \cite{katie} that implements the high energy factorization and provides an efficient way of computing off-shell gauge invariant amplitudes \cite{vanHameren:2012uj,vanHameren:2012if}. The collinear PDF describing the dilute projectile was set to CT18NLO and provided via LHAPDF \cite{Buckley:2014ana}. The small-$x$ dipole TMD gluon distribution was taken to be the Kutak-Sapeta (KS) fit to HERA data \cite{Kutak_2012}. This choice is dictated by the fact that its nonlinear evolution is based on the unified BK/DGLAP framework developed by Kwieciński and collaborators \cite{Kwiecinski:1997ee,Kutak:2003bd}, which includes, in particular, the kinematic constraint. The output from \KaTie\, are Les Houches (LHE) files which can then be processed by other Monte Carlo generators.

\item The second stage of the computation is to perform the parton shower and hadronization. As described in the Introduction, we use the  CASCADE Monte Carlo \cite{Baranov_2021}, which we run with the nonlinear TMD gluon distribution (with full awareness of the limitations of such a procedure). After performing the final-state and initial-state shower, taking into account the explicit transverse momentum of the off-shell gluons, CASCADE does the full hadronization utilizing routines from PYTHIA.

\end{enumerate}

In this work we shall be concerned with both the azimuthal correlations (defined to be the differential cross section in the azimuthal angle $\phi$ between the direct photon and a hadron) and the TEEC. 
The latter is defined as the transverse energy weighted differential cross-section and for photon-hadron production in p$A$ collisions can be written as \cite{gao2023}:
\begin{equation}
    \frac{d\Sigma^{\text{p} A \rightarrow \gamma \text{h}}}{d\phi} = \sum_{h} \int d \sigma^{\text{p}  A \rightarrow \gamma \text{h}} \frac{E_{T{\gamma}} E_{T{h}}}{E_{T{\gamma}} \sum_{h\prime} E_{T{h\prime}}} \delta ( \phi_{\gamma h} - \phi) \,
    \label{eqTEEC}
\end{equation}
where $E_{T{\gamma}}$ is the photon transverse energy, $E_{T{h}}$ the transverse energy of the hadron and $\phi_{\gamma h}$ is the azimuthal angle between the leading photon and each hadron, the sum being over all hadrons. Notice, that at parton level this is the same as the azimuthal correlation.

%%%%%%%%%%%%%%%%%%%%%%%%%%%%%%%%%%%%%%%%%%%%%%%%%%%%%%%%%%%%%

\section{Results}
\label{sec:Results}

In this section, we provide numerical results for TEEC and azimuthal correlations for $\gamma-$hadron production in proton-proton (pp) and proton-lead (p$Pb$) collisions within the FoCal detector acceptance as part of the future ALICE upgrade, as well as the extended ATLAS kinematics.

For the FoCal kinematics the parton-level events were generated at a center-of-mass energy of 8.8 TeV per nucleon in \KaTie, for both pp and p$Pb$. The colored partons were produced within an extended pseudorapidity window of $2.8 < \eta < 6.1$ (comparing to the experimental acceptance) and the direct photons with $3.8 < \eta < 5.1$. We further imposed a minimum transverse momentum condition of 7 GeV on both the parton and the direct photon (we discuss the sensitivity of this choice later in this section). For ATLAS kinematics the center-of-mass energy was set to 8.16 TeV per nucleon and the parton and the direct photon were required to have a minimum transverse momentum of 5 GeV. Here, the partons were produced within the extended pseudorapidity $1.7 < \eta < 5.0$, and the direct photons with $2.7 < \eta < 4.0$. Although the ATLAS detector covers both hemispheres, we only focus on the forward region for the purpose of this study. For both kinematical setups we require that the minimum distance $R$ between the partonic final states in the $\eta - \phi$ plane is $>0.1$. At parton level, the factorization and renormalization scales were set to the average $p_T$ of the final states. As mentioned above, the partons were generated within a slightly broader pseudorapidity window, as compared to the constraints applied on the hadrons. This is to account for hadrons produced within the detector acceptance but produced from partons emitted outside this range. 

\begin{table}[H] 
\centering  
    \begin{tabular}{p{1.2cm}lcc}
    \toprule
     & \textbf{Quantity} & \textbf{ATLAS} & \textbf{FoCal} \\
    \midrule
    \multirow{4}{*}{\rotatebox{90}{\shortstack{Parton \\ level}}} 
     & $p_{T_{\gamma}}$ & $> 5$ GeV & $> 7$ GeV \\
     & $p_{T_{ P}}$ & $> 5$ GeV & $> 7$ GeV \\
     & $\eta_{\gamma}$ & $2.7 < \eta_{\gamma} < 4.0$ & $3.8 < \eta_{\gamma} < 5.1$ \\
     & $\eta_{P}$ & $1.7 < \eta_{P} < 5.0$ & $2.8 < \eta_{P} < 6.1$ \\
    \midrule
    \multirow{5}{*}{\rotatebox{90}{\shortstack{Hadron \\ level}}} 
     & $p_{T_{\gamma}}$ & $> 5$ GeV & $> 5$ GeV \\
     & $p_{T_{h}}$ & ---  & ---  \\
     & $p_{T\, \text{Trig}}$ & $= 5\, (10)$ GeV & $= 5\, (10)$ GeV \\
     & $\eta_{\gamma}$ & $2.7 < \eta_{\gamma} < 4.0$ & $3.8 < \eta_{\gamma} < 5.1$ \\
     & $\eta_{h}$ & $2.7 < \eta_{h} < 4.0$ & $3.8 < \eta_{h} < 5.1$ \\
    \bottomrule
    \end{tabular}
    \caption{Parton and hadron level kinematical cuts for ATLAS and FoCal. The subscript $P$ corresponds to parton-level, while $h$ to hadron-level cut.}
    \label{kin}
\end{table}

\begin{figure}[hbt]
    \centering
    \includegraphics[width = \textwidth, height = 5.75cm]{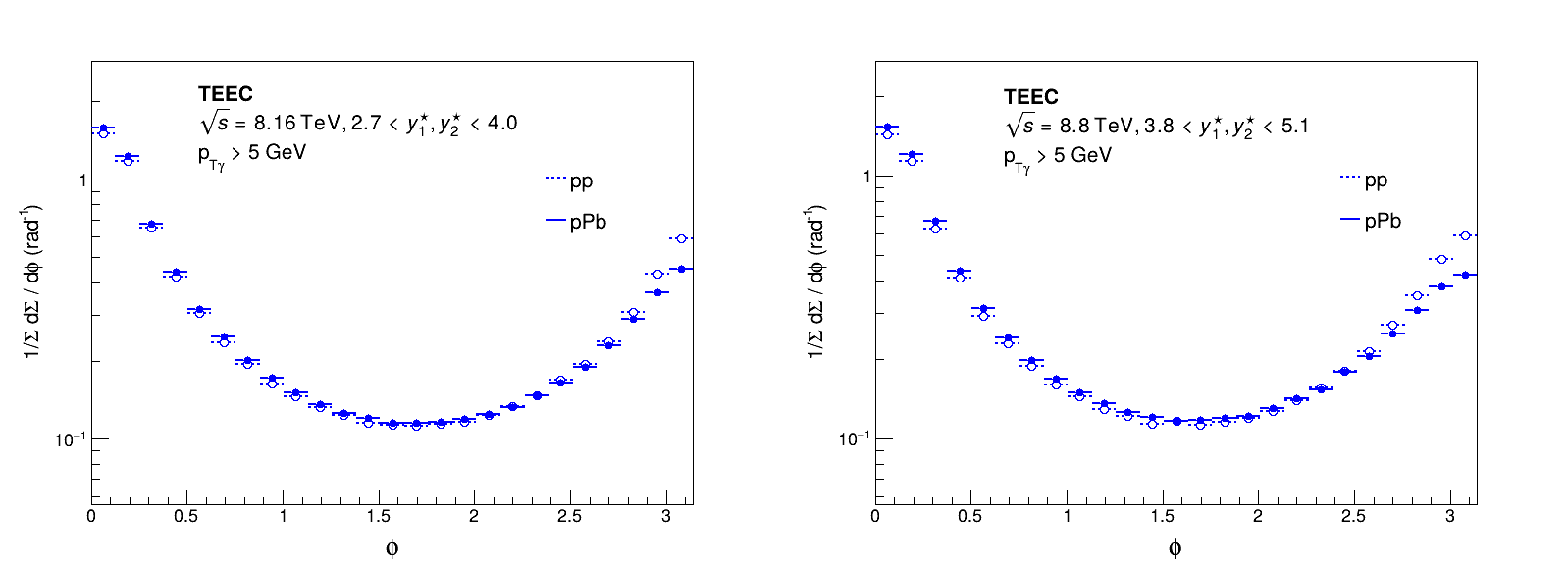}
    \caption{Normalized $\gamma$-hadron TEEC in pp and p$Pb$ collisions at CM energy 8.16 TeV, $2.7 < y_1^*, y_2^* < 4.0$ (on the left) and at CM energy 8.8 TeV, $3.8 < y_1^*, y_2^* < 5.1$ (on the right). The photon in both cases was required to have a minimum transverse momentum of 5 GeV. Although there was no such restriction on the hadron transverse momentum, the events were required to have at least one hadron with $p_T > 5$ GeV.}
    \label{TEEC}
\end{figure}

\begin{figure}[hbt]
    \centering
    \includegraphics[width = \textwidth, height = 5.5cm]{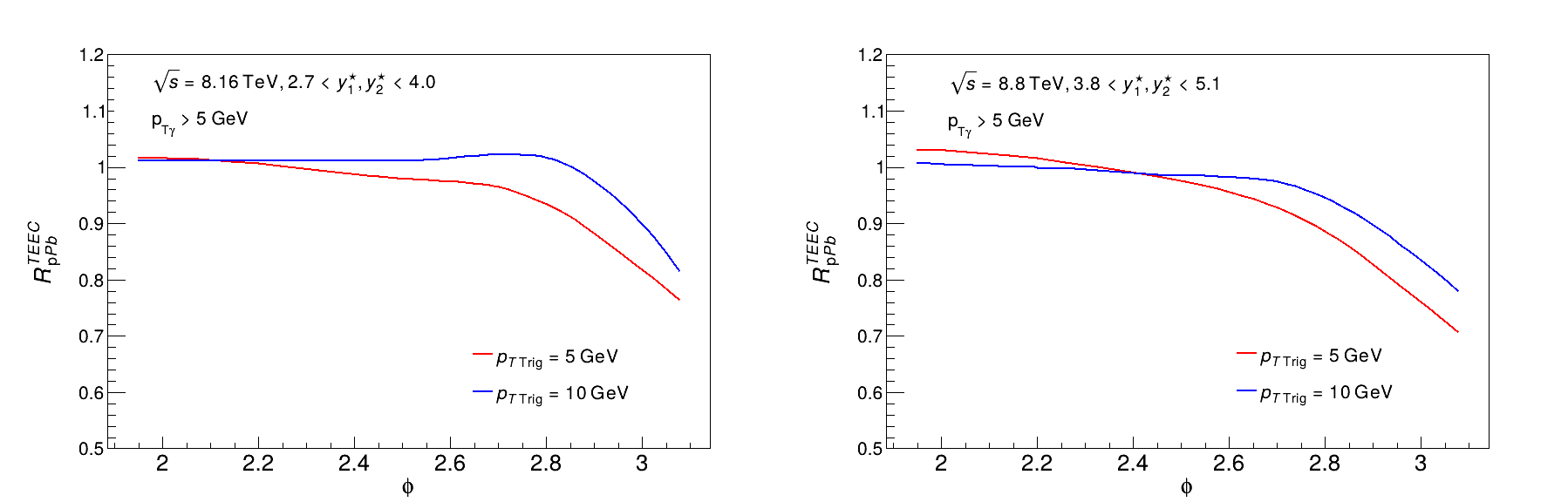}
    \caption{$R^{\mathrm{TEEC}}_{\mathrm{p}Pb}$ for $\gamma$-hadron production in pp and p$Pb$ collisions at CM energy 8.16 TeV, $2.7 < y_1^*, y_2^* < 4.0$ (on the left) and at CM energy 8.8 TeV, $3.8 < y_1^*, y_2^* < 5.1$ (on the right). The photon in both cases was required to have a minimum transverse momentum of 5 GeV. Although there was no such restriction on the hadron transverse momentum, the events were required to have at least one hadron with $p_T > 5$ GeV (in red) and $p_T > 10$ GeV (in blue).}
    \label{raTEEC}
\end{figure}

To construct the final observables we collect the photons and charged hadrons with $2.7 < \eta < 4.0$ for the ATLAS kinematics and with $3.8 < \eta < 5.1$ for FoCal. The photons were required to have a minimum transverse momentum of 5 GeV in both cases. No such transverse momentum constraints were imposed on the hadrons. However, we only choose the events where at least one hadron possesses transverse momentum above a $p_{T\,\mathrm{Trig}}$. We study the impact of this $p_{T\,\mathrm{Trig}}$ later in this section. We summarize these kinematic cuts in Table \ref{kin}:

Fig.~\ref{TEEC} shows $\gamma$-hadron TEEC in pp and p$Pb$ for both ATLAS (left) and FoCal (right) kinematics. There is a significant suppression in p$Pb$ TEEC compared to pp. The suppression is more pronounced in the back-to-back ($\phi \sim \pi$) region. Notice, that our TEEC is normalized, so the suppression is due to an interplay of the shape of the TMD distributions for proton and lead, and not just the overall suppression of the lead TMD. In these computations the $p_{T\,\mathrm{Trig}}$ was set to 5 GeV. In order to better visualize this suppression, in Fig.~\ref{raTEEC} we show the suppression factor $R^{\mathrm{TEEC}}_{\mathrm{p}Pb}$ defined as:
\begin{equation*}
    R^{\mathrm{TEEC}}_{\mathrm{p}Pb} = \frac{(\frac{1}{\Sigma}\frac{d \Sigma}{d \phi})_{\text{p}Pb}}{(\frac{1}{\Sigma}\frac{d \Sigma}{d \phi})_{\text{pp}}} \, .
\end{equation*}
We stress this is not the usual nuclear modification factor: (i) it is defined for the TEEC and (ii) both numerator and denominator are normalized quantities.
Although for a wide range of $\phi$ $R^{\mathrm{TEEC}}_{\mathrm{p}Pb}$ is close to 1, there is a sharp drop near the back-to-back region. We show $R^{\mathrm{TEEC}}_{\mathrm{p}Pb}$ for two different $p_{T\,\mathrm{Trig}}$ values, 5 GeV and 10 GeV. For a $p_{T\,\mathrm{Trig}}$ of 10 GeV (in blue), we see a suppression of about $\sim 20 \%$ for ATLAS kinematics (left) and about $\sim 22 \%$ for FoCal near $\phi \sim \pi$. For $p_{T\,\mathrm{Trig}} = 5$ GeV (in red), we see a stronger suppression of about $\sim 24 \%$ for ATLAS (left) and about $\sim 30 \%$ for FoCal kinematics near $\phi \sim \pi$. This shows how the shape of the distribution varies with the $p_{T\,\mathrm{Trig}}$.

\begin{figure}[hbt]
    \centering
    \includegraphics[width = \textwidth, height = 5.75cm]{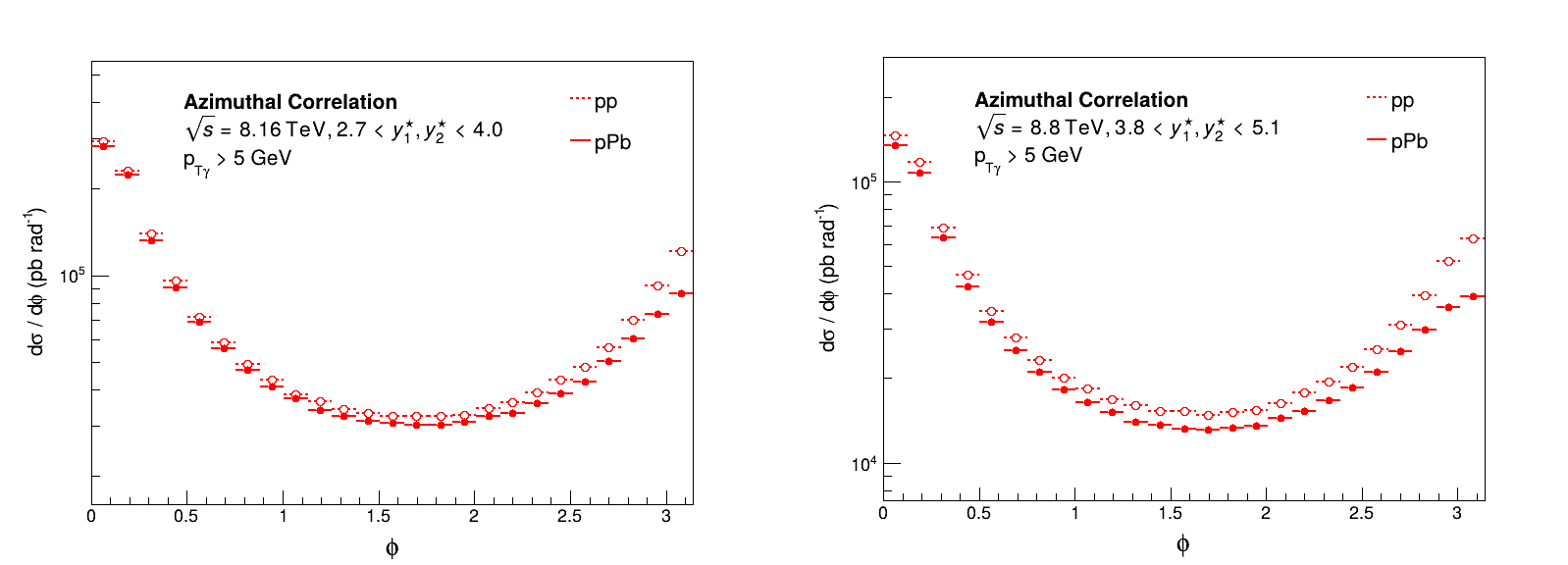}
    \caption{Leading $\gamma$-hadron azimuthal correlation in pp and p$Pb$ collisions at CM energy 8.16 TeV, $2.7 < y_1^*, y_2^* < 4.0$ (on the left) and at CM energy 8.8 TeV, $3.8 < y_1^*, y_2^* < 5.1$ (on the right). The photon in both cases was required to have a minimum transverse momentum of 5 GeV. Although there was no such restriction on the hadron transverse momentum, the events were required to have at least one hadron with $p_T > 5$ GeV.}
    \label{leadingAC}
\end{figure}

\begin{figure}[hbt]
    \centering
    \includegraphics[width = \textwidth, height = 5.5cm]{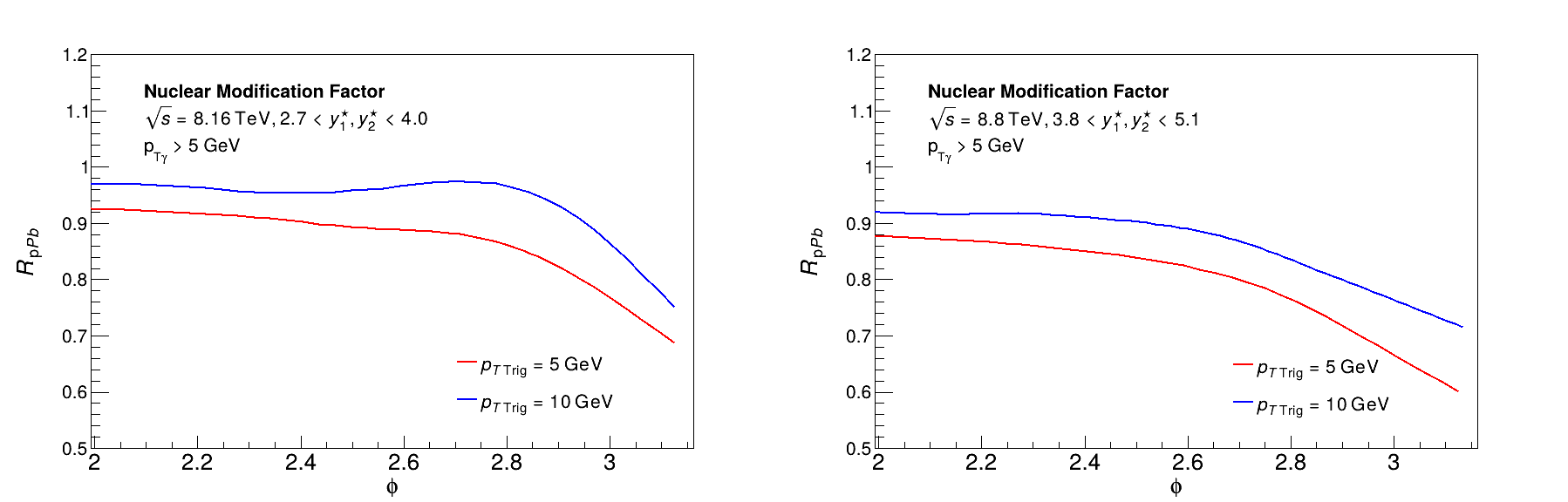}
    \caption{Nuclear modification factor for $\gamma$-hadron production in pp and p$Pb$ collisions at CM energy 8.16 TeV, $2.7 < y_1^*, y_2^* < 4.0$ (on the left) and at CM energy 8.8 TeV, $3.8 < y_1^*, y_2^* < 5.1$ (on the right). The photon in both cases was required to have a minimum transverse momentum of 5 GeV. Although there was no such restriction on the hadron transverse momentum, the events were required to have at least one hadron with $p_T > 5$ GeV (in red) and $p_T > 10$ GeV (in blue).}
    \label{ra}
\end{figure}

In Fig.~\ref{leadingAC} we show the azimuthal correlation (defined as the differential cross section in the azimuthal angle) of the leading photon-hadron pair within the ATLAS (left) and FoCal (right) kinematics. $p_{T\,\mathrm{Trig}}$ was taken to be 5 GeV. Similar to the TEEC distributions, we observe a significant suppression in p$Pb$ differential cross-section compared to pp, which is much stronger in the back-to-back region. This is consistent with the CGC picture, which predicts a suppression of the lead TMD (normalized to the number of nucleons) at small-$x$ comparing to the proton. The results in the FoCal kinematics correspond to a stronger suppression due to more forward kinematics. Let us note, that, in general, it is not obvious that the suppression due to gluon saturation survives the shower and hadronization. For the present kinematics our computation shows, however, that the suppression is  still present and is rather significant.  The usual nuclear modification factor $R_{\mathrm{p}Pb}$ defined as:
\begin{equation*}
    R_{\mathrm{p}Pb} = \frac{1}{A} \frac{(\frac{d \sigma}{d \phi})_{\text{p}Pb}}{(\frac{d \sigma}{d \phi})_{\text{pp}}} \, ,
\end{equation*}
is shown in Fig.~\ref{ra}. Here too, we show the nuclear modification factor for two different $p_{T\,\mathrm{Trig}}$ values of 5 and 10 GeV. $R_{\mathrm{p}Pb}$ exhibits a stronger suppression for both $p_{T\,\mathrm{Trig}}$ values in comparison to the TEEC ratio, especially in the back-to-back region. The factor drops by $\sim 30\%$ for $p_{T\,\mathrm{Trig}}$ 5 GeV (in red) for ATLAS (left) and close to $\sim 40\%$ for FoCal (right) kinematics. Similarly for $p_{T\,\mathrm{Trig}} = 10$ GeV, there's a stronger suppression of $\sim 30\%$ observed in FoCal, as compared to $\sim 25\%$ in ATLAS.

\begin{figure}[hbt]
    \centering
    \includegraphics[width = \textwidth, height = 5.75cm]{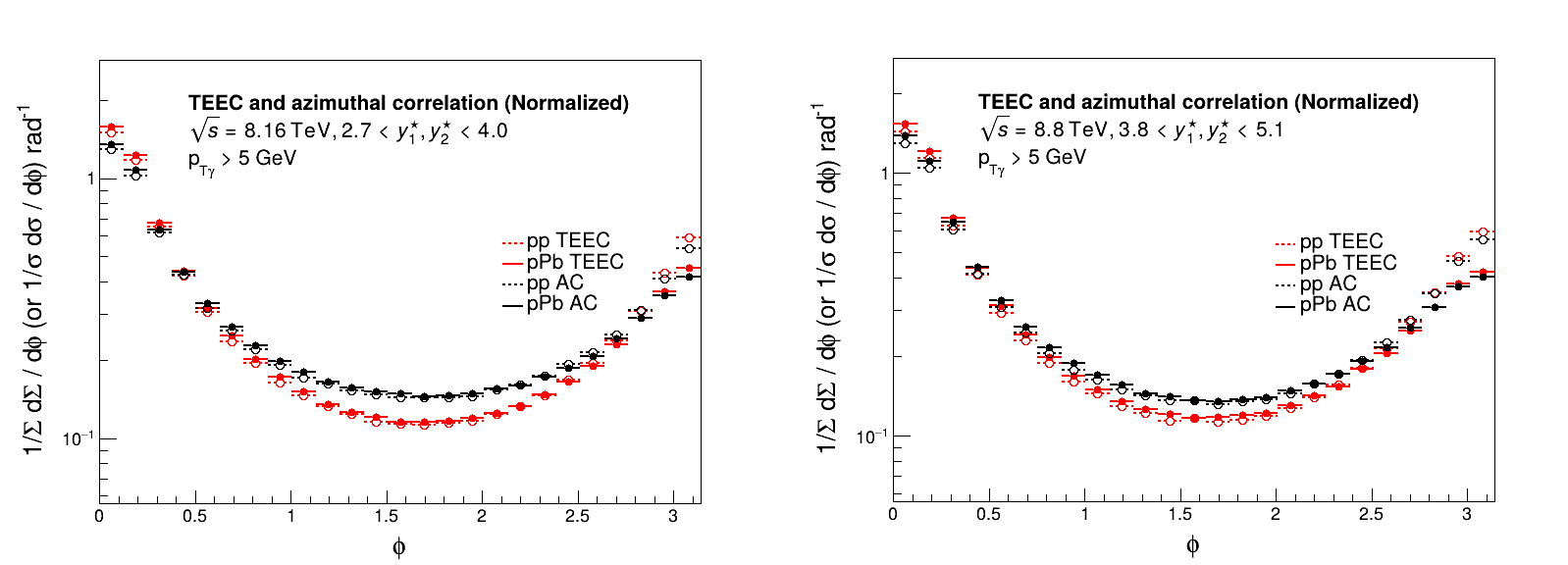}
    \caption{Comparison of normalized TEEC and azimuthal correlation of the hardest $\gamma$-hadron pair in pp and p$Pb$ collisions at CM Energy 8.16 TeV, $2.7 < y_1^*, y_2^* < 4.0$ (left) and at CM energy 8.8 TeV, $3.8 < y_1^*, y_2^* < 5.1$ (right). The photon in both cases was required to have a minimum transverse momentum of 5 GeV. Although there was no such restriction on the hadron transverse momentum, the events were required to have at least one hadron with $p_T > 5$ GeV.}
    \label{leadhAC}
\end{figure}

\begin{figure}[hbt]
    \centering
    \includegraphics[width = \textwidth, height = 5.75cm]{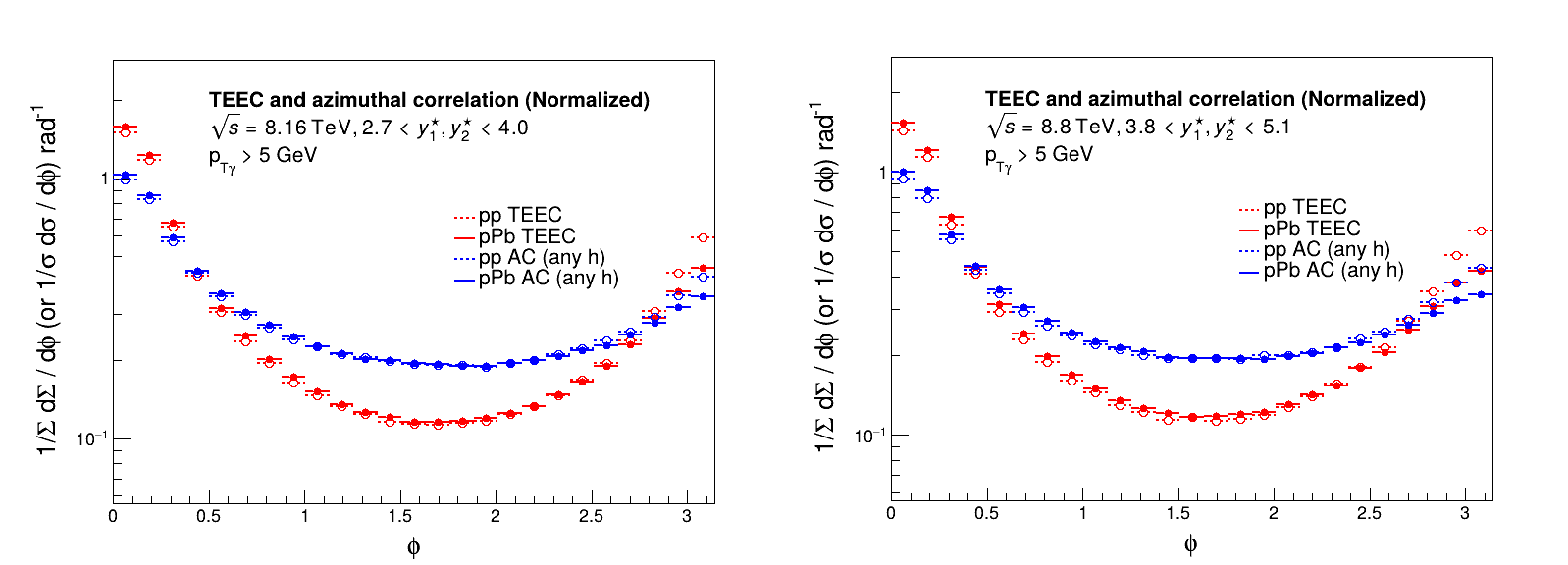}
    \caption{Comparison of normalized TEEC and azimuthal correlation of the hardest $\gamma$ and a randomly chosen hadron in pp and p$Pb$ collisions at CM Energy 8.16 TeV, $2.7 < y_1^*, y_2^* < 4.0$ (left) and at CM energy 8.8 TeV, $3.8 < y_1^*, y_2^* < 5.1$ (right). The photon in both cases was required to have a minimum transverse momentum of 5 GeV. Although there was no such restriction on the hadron transverse momentum, the events were required to have at least one hadron with $p_T > 5$ GeV.}
    \label{anyhAC}
\end{figure}

In our study, we defined the azimuthal correlation observable for the leading photon and the leading hadron. Alternatively, we could have defined this observable for the leading photon and \emph{any} hadron (with the presence of a trigger hadron in both cases). Let us therefore discuss how these hadron-level observables compare with each other and with the TEEC. Fig.~\ref{leadhAC} shows the TEEC and the azimuthal correlations of the leading photon-hadron pair and Fig.~\ref{anyhAC}, on the other hand, is the azimuthal correlation of the leading photon with a randomly chosen hadron. The azimuthal correlation distributions have been normalized for comparison purposes. Although the overall shape is quite similar, the normalized azimuthal correlations are generally flatter, especially the one defined for leading-photon--any-hadron correlation. This picture is consistent with our expectations from perturbative QCD, namely the selection of the leading hadron corresponds to the harder partonic process, which is naturally very peaked in both collinear and back-to-back regions.

\begin{figure}[hbt]
    \centering
    \includegraphics[width = \textwidth, height = 5.5cm]{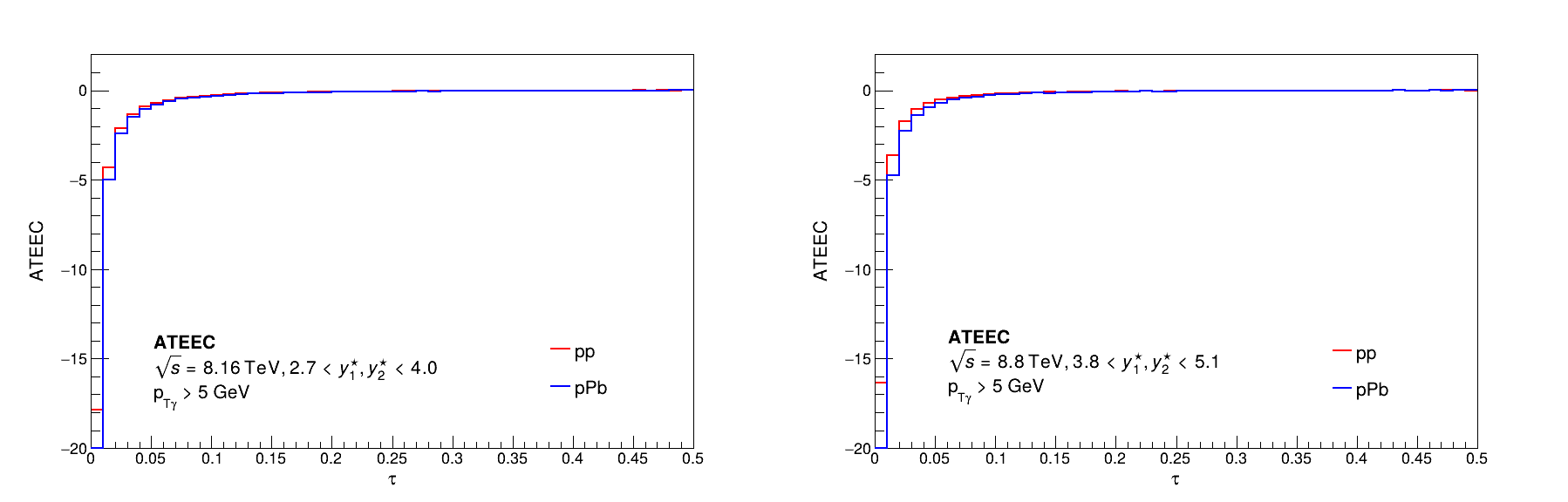}
    \caption{Asymmetry in $\gamma$-hadron TEEC (ATEEC) in pp and p$Pb$ collisions at CM energy 8.16 TeV, $2.7 < y_1^*, y_2^* < 4.0$ (on the left) and at CM energy 8.8 TeV, $3.8 < y_1^*, y_2^* < 5.1$ (on the right). The photon in both cases was required to have a minimum transverse momentum of 5 GeV. Although there was no such restriction on the hadron transverse momentum, the events were required to have at least one hadron with $p_T > 5$ GeV.}
    \label{ateec}
\end{figure}

Let us discuss next the 
 asymmetry of TEEC (ATEEC), which is defined as the difference: 
\begin{equation} \label{eqAT}
    \textbf{ATEEC} = \frac{1}{\Sigma} \frac{d \Sigma}{d \tau} |_{\tau} - \frac{1}{\Sigma} \frac{d \Sigma}{d \tau} |_{1 - \tau} \, ,
\end{equation}
where $\tau = \frac{1\,+ \,\text{cos}\, \phi}{2}$. ATEEC has been found to be quite useful in analyzing the shape of TEEC distributions and in practice has been used alongside the TEEC. Fig.~\ref{ateec} shows this asymmetry for ATLAS (left) and FoCal (right) kinematics. The quantity is negative throughout. It is highly asymmetric for both pp and p$Pb$ in the low $\tau$ ($\tau \sim 0$) region, which corresponds to the back-to-back limit. It appears that p$Pb$ TEEC is more asymmetric than pp, although both sets of distributions become more symmetric towards $\tau \sim 0.5$ or $\phi \sim \pi/2$. 

\begin{figure}[H]
    \centering
    \includegraphics[width=0.85\textwidth, height = 10cm]{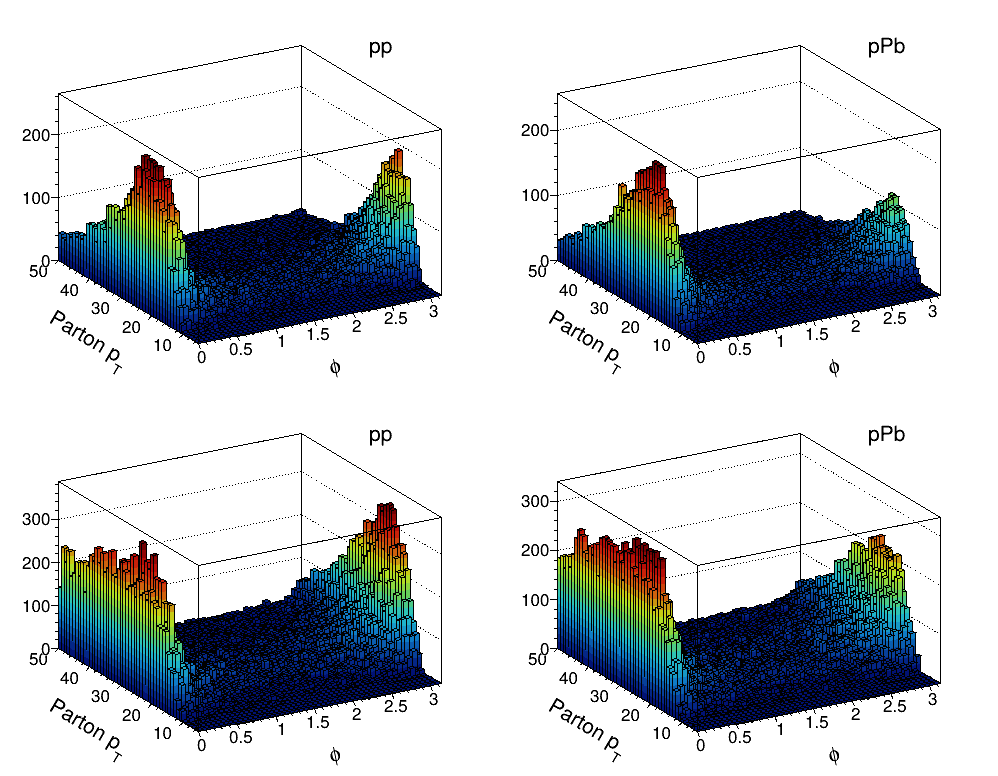}
    \caption{$\gamma$-hadron TEEC counts in pp and p$Pb$ collisions at CM Energy 8.16 TeV, $2.7 < y_1^*, y_2^* < 4.0$ (bottom row) and CM energy 8.8 TeV, $3.8 < y_1^*, y_2^* < 5.1$ (top row) w.r.t. their azimuthal separation ($\phi$) and the $p_T$ of hardest parton. The vertical axis here reflects event counts, and not the TEEC defined above.}
    \label{comb2D}
\end{figure}

Let us finally see how much does the earlier assumption about the minimum parton transverse momentum impact the final observables. Here, we mean the cut imposed in \KaTie\, at the parton-level of the computation. Fig.~\ref{comb2D} shows a 2D histogram, where the z-axis represents event counts with respect to the azimuthal angle between the leading photon and each hadron accounted, with the corresponding $p_T$ of the leading parton (in the partonic subprocess). For this objective, the trigger was set to 10 GeV. As seen from the plots, the cuts imposed on hadrons are sufficient to induce the proper parton-level kinematics. In particular, there is very little contribution from events with partons having $p_T$ smaller than the trigger.
\begin{figure}[hbt]
    \centering 
    \includegraphics[width=0.55\textwidth, height = 7cm]{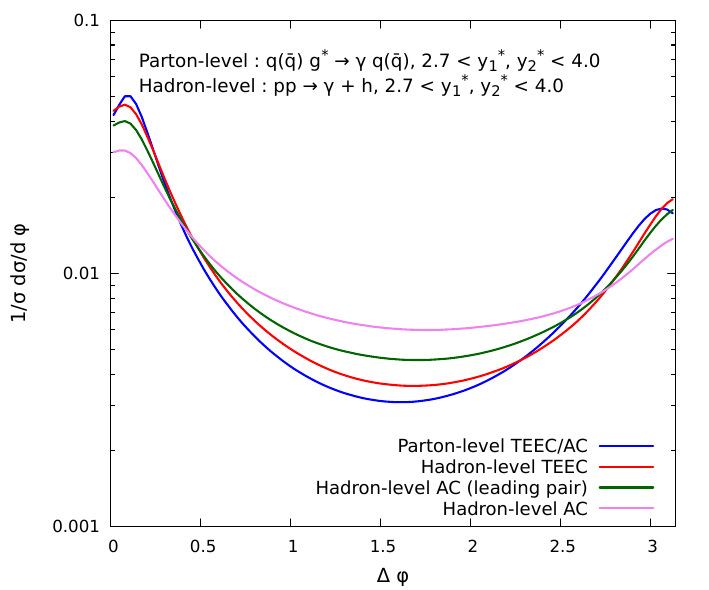}
    \caption{Comparison of parton level azimuthal correlation from \KaTie, with $\gamma$-hadron TEEC and azimuthal correlation as obtained post hadronization via CASCADE in pp collisions at CM energy 8.16 TeV, with $2.7 < y_1^*, y_2^* < 4.0$ (all normalized). The partons were generated with a minimum transverse momentum of 5 GeV in \KaTie. }
    \label{compare}
\end{figure}
Finally, Fig.~\ref{compare} shows the comparison of these observables with the parton-level azimuthal correlations (we plot this only for $pp \rightarrow \gamma + h$ in ATLAS kinematics for brevity). More precisely, we compare the parton-level correlation with the hadron-level TEEC (red) and azimuthal correlation of both the leading pair (in green) and leading photon with a randomly chosen hadron (in pink). Obviously, at parton-level the TEEC is identical with azimuthal correlation between the direct photon and the parton; both the direct photon and parton were required to have a minimum transverse momentum of 5 GeV and were generated within $2.7 < \eta < 4.0$ in \KaTie. The hadron-level TEEC and azimuthal correlations were computed as explained earlier. As expected, the parton-level correlations (or TEEC) are very peaked in the collinear and back-to-back regions. The TEEC, despite being a hadron-level observable,  quite closely follows the parton dynamics.

\section{Summary}
\label{sec:Summary}

In this work, we presented detailed numerical results for forward $\gamma$-hadron TEEC and azimuthal correlations in both proton-proton (pp) and proton-lead (p$Pb$) collisions within the kinematics of the near-future ALICE and ATLAS upgrades. We used the CGC-based formalism in the form of the hybrid $k_T$ factorization, with the TMD gluon distributions encoding the gluon saturation effects at small-$x$. We used a two-stage procedure to obtain the hadron-level results. The parton-level computation was done using the \KaTie\, Monte Carlo, while the hadronization and parton shower was performed using CASCADE. It is worth mentioning at this point, that although this computation incorporates the nonlinear gluon dynamics, there is no explicit parton-recombination implemented in the CASCADE. On the other hand, the rate of parton production in the shower is correctly taken into account. We therefore claim that such a computation captures the most important features. 

We observe a significant suppression of both the normalized TEEC and azimuthal correlations in p$Pb$ compared to pp. In the former, the suppression is due to a modification of the $k_T$ dependence of the lead TMD, comparing to the proton TMD (the nuclear broadening). In the latter case, the suppression is much stronger and is mainly due to the suppression of the maximum of the dipole lead TMD, comparing to the proton TMD at small $x$. The observed trends align with predictions based on gluon saturation. Our study further highlights the potential of future forward detectors at LHC. In particular, the FoCal calorimeter, with its very forward pseudorapidity coverage, probes the nuclear target at significantly smaller values of $x$, making it particularly sensitive to the potential gluon saturation signals.

\section*{Acknowledgments}
We thank I. Grabowska-Bołd for discussions. The work was suported by the Polish National Science Centre, grant no. 2020/39/O/ST2/03011.

\bibliographystyle{JHEP} 
\bibliography{references,references1,references2}

\end{document}